\documentclass{appolb}
\usepackage{mxedruli}

\begin{document}

\title{Breaking the light speed barrier}

\author{ O.~I.~Chashchina
\address{\'{E}cole Polytechnique, Palaiseau, France and
Department of physics, Novosibirsk State University, 630 090,
Novosibirsk, Russia}\\ \vspace*{3mm}
Z.~K.~Silagadze
\address{Budker Institute of Nuclear Physics SB RAS and Novosibirsk 
State University, 630 090, Novosibirsk, Russia}
}


\maketitle

\begin{abstract}
As it is well known, classical special relativity allows the existence of three
different kinds of particles: bradyons, luxons and tachyons. Bradyons have
non-zero mass and hence always travel slower than light. Luxons are particles
with zero mass, like the photon, and they always travel with invariant
velocity. Tachyons are hypothetical superluminal particles that always move
faster than light. The existence of bradyons and luxons is firmly established,
while the tachyons were never reliably observed. In quantum field theory, the
appearance of tachyonic degrees of freedom indicates vacuum instability rather
than a real existence of the faster-than-light particles. However, recent
controversial claims of the OPERA experiment about superluminal neutrinos
triggered a renewed interest in superluminal particles. Driven by a striking
analogy of the old Frenkel-Kontorova model of a dislocation dynamics to the
theory of relativity, we conjecture in this note a remarkable
possibility of existence of the forth type of particles, elvisebrions, which
can be superluminal. The characteristic feature of elvisebrions, distinguishing
them from tachyons, is that they are outside the realm of
special relativity and their energy remains finite (or may even turn to zero)
when the elvisebrion velocity approaches the light velocity.
\end{abstract}
\PACS{03.30.+p, 11.30.Cp, 14.80.-j}

\section{Introduction}
Superluminal sources of radiation were first considered by Heaviside in
1888 and in the following years he derived most of the formalism
of what is nowadays called Cherenkov radiation \cite{1,2,3}. Sommerfeld,
being unaware of Heaviside's insights, also considered electromagnetic
radiation from superluminal electrons \cite{1,4}. However, the timing when 
these works occured was unfortunate \cite{5} because Einstein's first paper on
special relativity has appeared a few months after Sommerfeld's
1905 publication on superluminal electrons and it became clear that electrons
and all other particles with nonzero mass cannot be accelerated beyond the
light velocity in vacuum. As a result we had to wait for several decades
before the accidental, experimental discovery of the Cherenkov radiation in
1934 \cite{6} and even more so to realize that special relativity does not
prohibit superluminal sources of radiation \cite{7}.

Of course, these superluminal sources of radiation cannot be individual
electrons or other Standard Model charged particles which are ordinary
bradyons and hence cannot overcome the light-speed barrier. Nothing
precludes though the aggregates of such particles to produce superluminally
moving patterns in a coordinated motion \cite{7,8}. The simplest example 
of such a superluminally moving pattern is a light spot produced by a rotating 
source of light on a sufficiently remote screen. One can imagine a three 
dimensional analog of such a superluminal light spot, namely a radiation pulse 
with a conical frontal surface as a result of light reflection by a conical 
mirror. The vertex of this conical frontal surface is a focus which can travel
superluminally and the field energy density at this spot is several orders
of magnitude higher than in a flat light spot, making this object look like a
particle \cite{9}.

One may argue that the light spot is not a real object and its propagation
in space is not a real process at all since it does not transfer an
energy from one point to another on its path \cite{10}. However, already in
classical physics it is not easy to give a general definition of what a real
thing is, without even speaking about the quantum theory \cite{11}. As 
a result, our understanding of what kind of velocities are limited by special
relativity continues to evolve \cite{12,13}.

Recently the OPERA experiment reported an alleged evidence for superluminal
muon neutrinos \cite{14}. Although it was evident from the beginning that this 
experimental result contradicts all that we know about neutrinos and weak 
interactions \cite{15,16,17,18}, and hence was most probably due to some 
unaccounted systematic errors \cite{19,20,21,22,23}, it has generated a huge 
interest in our postmodern physics community. Many explanations of this 
unexpected and surprising result, one more fantastic than another, were  
proposed in literature. We cite only a few representatives which are 
potentially interesting but in our opinion improbable  \cite{24,25,26,27,28,
29,30}.

The Lorentz invariance is one of the most experimentally well established and 
tested feature of Nature \cite{31,32,33,34}. In light of this impressive 
experimental evidence, it is not surprising that finally no indications of 
superluminality in the neutrino sector were found out, as expected, and it 
seems the original anomaly was most probably due to the OPERA equipment
malfunctioning \cite{77,78,79}.

However, ``There are more things in Heaven and Earth, Horatio, than are
dreamt of in your philosophy'' \cite{35}, and we cannot be ``certain that
Nature has exhausted her bag of performable tricks'' \cite{36}. Therefore,
it makes sense to ask in a broader context whether the established 
unprecedented high accuracy of Lorentz invariance precludes a superluminal 
energy transfer at moderate energy scales in all conceivable situations. 
As we will try to argue in this paper, the answer is negative.

\section{Tachyons}
In 1905, Einstein published his paper on special relativity \cite{36A} in
which he concluded that ``speeds in excess of light have no possibility of
existence''. For many years this has become an axiomatic statement, and any
assumptions that were contrary to this dogma were perceived with a bias, as
unscientific fantasies.

The reason behind the Einstein's conclusion was that according to the theory
of relativity you need an infinite amount of energy to accelerate a particle
to the speed of light. Also, the special relativistic relationship between
particle's energy and its mass implies that the mass of a particle moving
with velocity  $v>c$ would be imaginary and hence "unphysical". This is also
applied to other physical quantities, such as the proper time and the proper
length. Finally, it was believed that if such particles exist, the principle
of causality would be violated as they can be used to send information in
the past (the so called Tolman antitelephone paradox \cite{36B}).

Interestingly, despite being a proponent of the concept of a
velocity-dependent electromagnetic mass, Heaviside never acknowledged this
limitation on the particle's velocity \cite{1}, and maybe for a good reason.
In fact, Einstein's conclusion is fallacious, even absurd. As eloquently
expressed by Sudarshan in 1972, this is the same as asserting ``that there
are no people North of the Himalayas, since none could climb over the mountain
ranges. That would be an absurd conclusion. People of central Asia are born
there and live there: they did not have to be born in India and cross the
mountain range. So with faster-than-light particles'' \cite{36C}.

Probably, Einstein was well aware of the weakness of the infinite energy
argument. In fact, Tolman's antitelephone paradox was invented by him
\cite{36D}, and it is indeed a serious conundrum and basic problem for any
theory involving faster-than-light propagation of particles. Its essence is
the following.

For events separated by a spacelike interval, their relative time order is
not invariant but depends on the choice of reference frame. However, the
interval between the emission and absorption events of a superluminal
particle is just spacelike. Therefore, in some inertial reference frames
the superluminal particle will be absorbed before it is emitted, and it
appears that we have a grave problem with causality.

However, the same problem is already present in quantum field theory that
unifies the fundamental ideas of special relativity and quantum mechanics and
conforms the modern basis of elementary particle physics. In quantum field
theory, the amplitude for a particle to propagate from a space-time point
$x=(x_0,\vec{x})$ to a point $y=(y_0,\vec{y})$ is Lorentz invariant and is
given by the Wightman propagator ($\hbar=c=1$ is assumed) \cite{36E}
\begin{equation}
D(x-y)=\int\frac{d\vec{p}}{(2\pi)^3}\,\frac{e^{-ip\cdot(x-y)}}{2\sqrt{
\vec{p}^{\,2}+m^2}}.
\label{eq1-1}
\end{equation}
When the difference $x-y=(0,\vec{r})$ is purely in the spatial direction, the 
integral (\ref{eq1-1}) can be evaluated by:
\begin{equation}
D(x-y)=-\frac{1}{4\pi^2 r}\,\frac{\partial}{\partial r}\int\limits_0^\infty
\frac{\cos{(pr)}}{\sqrt{p^2+m^2}}=-\frac{1}{4\pi^2 r}\,\frac{\partial}
{\partial r}K_0(mr)=\frac{m}{4\pi^2 r}K_1(mr),
\label{eq1-2}
\end{equation}
where $K_0$ and $K_1$ are modified Bessel functions of the second kind. 
Using the well known asymptotics
$$K_1(x)\approx \sqrt{\frac{\pi}{2x}}\,e^{-x},\;\;\;\mathrm{if}\;\;\;x\ll 1,$$
we see that within its Compton wavelength, $m^{-1}$, a particle has a
significant probability to propagate with infinite velocity (with respect to
this particular reference frame in which $x-y=(0,\vec{r})$).

For particles of a very small mass (neutrinos) the Compton wavelength can be
macroscopically large. Interestingly, this kind of superluminal propagation
of neutrinos within their Compton wavelength was even suggested as a possible
explanation of the OPERA anomaly (then not yet falsified) \cite{30,36F}, 
but shown to be non-working \cite{36F}.

Quantum field theory offers a miraculously clever solution of this
superluminal propagation dilemma \cite{36E,37A,37B}. Suppose that in the 
reference frame $S$ a particle propagates superluminally between the points 
$x$ and $y$ separated by spacelike interval $(x-y)^2<0$, and suppose that 
$x$ is the emission point and $y$ is the absorption point so that 
$x_0<y_0$. Since the
interval is spacelike, there exist another reference frame $S^\prime$ such
that $x^\prime_0>y^\prime_0$ and, therefore, in this frame the particle
propagates backward in time: its absorption precedes its emission in
apparent violation of causality. However, in the frame $S^\prime$ the
particle's energy is negative as it can be easily checked using the
Lorentz transformation properties of the energy-momentum four-vector.
But a negative energy particle propagating backward in time is nothing more 
than a positive energy antiparticle propagating forward in time. This
Feynman-Stueckelberg interpretation of antiparticles is at the heart of
quantum field theory's resolution of superluminal propagation dilemma.
The observer in the frame $S^\prime$ does not see that the particle is
absorbed at $y$ before its emission at $x$, instead he/she sees the
antiparticle emitted at $y$ and absorbed at $x$, therefore he/she has no
apparent reason to worry about causality violation.

On a deeper level, for causality to be restored one needs not to suppress
a superluminal propagation of particles but to ensure that any measurement
(disturbance) at a space-time point $x$ cannot influence an outcome of
another measurement at space-time point $y$ if the points are separated by
a spacelike interval. Evoking antiparticles, quantum field theory ensures the
cancellation of all acausal terms in commutators of two local observables
at spacelike separation and  does not allow information to be transmitted
faster than the speed of light.

Many subtleties and open questions remain, however, because it is not 
a trivial task to merge quantum mechanics, with its notorious non-localities, 
and special relativity \cite{37C,37D,37E,37F}. ``Relativistic causality -
formulate it as you like! - is a subtle matter in relativistic quantum
theories'' \cite{37G}. We just mention two interesting examples where the
alleged superluminal effects can be interpreted as being due to the
propagation of virtual photons outside of the light cone. Nevertheless, no 
one of them allows messages to be transmitted faster than the speed of light.

It can be shown that entanglement and mutual correlations can be generated at
space-like separated points \cite{37F}. Of course, this problem is as old
as the Einstein-Podolsky-Rosen paradox \cite{37-EPR}.

Another example is the so called Hartman effect. Quantum mechanics predicts
that the transmission time across a potential barrier becomes independent of
barrier thickness for very thick barriers \cite{37-HRTM}. This strange
prediction was experimentally confirmed in frustrated total internal
reflection, which is an optical analog of quantum mechanical tunneling
\cite{37-FTIR,37-FTIR1}, and in other optical tunneling experiments
\cite{37-NIMTZ}. Apparent superluminal behavior in such experiments is
related to evanescent modes, a kind of classical analog of virtual photons
\cite{37-NIMTZ1}.

How real are virtual photons? Sometimes virtual particles are considered as
pure mathematical constructions, just a tool to visualize perturbation theory
calculations. However, there are many things in modern physics which can not
be observed as separate asymptotic states and nevertheless nobody questions
their real existence, quarks being the most notorious example. Another
example is short-lived particles, like $\omega$ and $\phi$ mesons. Therefore,
we cannot deny a kind of existence of virtual particles and hence of the
superluminal phenomena associated with them.

Anyway it seems we have no compelling reason from special relativity against
tachyons, alleged superluminal particles, and it is surprising that the first 
serious papers on tachyons appear only in the early sixties of the past 
century. In 1962, Sudarshan, Bilaniuk and Deshpande, not without the
help of personal contacts, published their article ``Meta relativity''
\cite{37}, which became the starting point of serious thinking about tachyons.
Fast enough, this article became famous and has induced many debates and other
publications (see \cite{38,39,40} and references therein). In these 
publications it was discussed whether the existence of tachyons is consistent 
with the theory of relativity and also the formalism for quantum theory of 
tachyons was developed. The term ``tachyon'' itself (from the Greek 
$\tau\alpha\chi\upsilon\varsigma$, meaning  ``swift'') was proposed by Gerald 
Feinberg in 1967 for particles with a velocity greater than the speed of 
light \cite{38}.

According to these studies, tachyons, bradyons and luxons constitute three
independent groups of particles that cannot be converted into each other by
Lorentz transformations. Thus, all particles that move relative to us with a
speed lower than the speed of light we perceive as bradyons. When
accelerating, the velocity of a bradyon increases up to the speed of light
but even despite the consumption of any finite amount of energy, never 
reaches it. Tachyons have
their superluminal velocities not due to acceleration but because they are
born with $v>c$ velocities, like photons (luxons) are always born with
velocity $v=c$. With respect to any system of bradyon observers, tachyons
always travel at a speed greater than the speed of light. There is
no reference frame, equivalent to our own frame up to a Lorentz
transformation, which would be the rest frame for a tachyon, so even in 
principle, we are not
able to make measurements of its mass or proper length.
According to the equations of special relativity, the mass and proper length 
of a tachyon turn out to be imaginary, but this does not contradict the 
principle that all observable physical quantities must be real, because 
finally we are not able to measure these quantities, and so they are 
unobservable.

The principle of causality is also not violated by tachyons much in the same 
way as it is not violated in quantum field theory thanks to the
Feynman-Stueckelberg interpretation of antiparticles. We can conclude 
then that special relativity does not prohibit tachyons and therefore,
they must exist according to  the Gell-Mann's totalitarian principle 
``everything not forbidden is compulsory'' \cite{39} (in fact, this wonderful 
phrase first appeared in T.~H.~White's fantasy novel {\it The Once and Future 
King} \cite{40A}. Sometimes the phrase is erroneously attributed to 
George Orwell's famous novel {\it Nineteen Eighty-Four}, see for example 
\cite{40B}. We were unable to find the phrase in the Orwell's novel). 

Tachyons were searched but never reliably found \cite{40,40C}.
Although there are some observed anomalies in extensive air showers which
could be attributed to tachyons \cite{40C}, the evidence is not conclusive 
enough. It seems that the Gell-Mann's totalitarian principle fails for 
tachyons, but why?

The clue for the resolution of this enigma is to realize that the 
totalitarian principle is about quantum theory and ``the break that quantum 
mechanics introduces in the basic underlying principles that have been 
working through history in the human thought since immemorial times, is 
absolute'' \cite{40D}. The truth is that the Gell-Mann's totalitarian 
principle does not fail at all and tachyons do exist. However the meaning of 
``exist'' is quit different from what is usually assumed.

First of all, tachyons exist as virtual particles. In fact, every elementary
particle can become tachyonic as a virtual particle. Note that up to now
we have emphasized superluminality as a defining property of tachyons. This 
is justified when we are talking about tachyons in the framework of special
relativity, because special relativity is essentially a classical theory,
but is no longer justified in quantum theory with its radical distinction from
classical concepts. For example, when the evanescent modes in the photon
tunneling experiments are considered as virtual photons and claimed that
they propagate superluminally, this is not quite correct. Classical concept of 
propagation velocity is not well-defined for evanescent modes or virtual 
photons. Nothing well defined and localized propagates through the tunneling 
barrier passing continuously through every point along the trajectory.

The notion of particles, which we have borrowed from the classical physics,
is also not quite satisfactory. Instead of talking about dubious 
wave-particle duality, which is a concept as incoherent \cite{40E} as 
the devil's pitchfork, a classic impossible figure \cite{40F}, it is better 
to accept from the beginning that the objects that we call elementary 
particles are neither particles nor waves but quantons, some queer objects 
of the quantum world \cite{41A,41B}.

The best way to classify elementary quantons is the use of space-time 
symmetry, where the elementary quantons correspond to the irreducible unitary 
representations of the Poincar\'{e} group \cite{41C,41D,41E}, first given by
Wigner \cite{41C}. The norm of the energy-momentum four-vector, 
$P_\mu P^\mu=m^2$, is a Casimir invariant of the Poincar\'{e} group and hence
its value partially characterizes a given irreducible representation. If 
$m^2>0$, positive energy representations are classified by the mass $m$ and 
the spin $s$ which comes from the compact stabilizer subgroup $SO(3)$ (or, 
better, from its double cover $SU(2)$). 

In the massless case, $m=0$, 
irreducible representations of the Poincar\'{e} group are induced by the 
Euclidean stabilizer subgroup $E(2)$ which is non-compact and has no 
finite-dimensional representations other than trivial. The trivial 
one-dimensional representation of $E(2)$ induces the irreducible 
representations of the Poincar\'{e} group, labeled by the helicity, 
describing photons and other massless particles. 

Usually one discards
irreducible representations of the Poincar\'{e} group induced by 
infinite-dimensional representations of $E(2)$ (the so called continuous 
spin representations) because the corresponding particles have been never 
experimentally observed, ``but there is no conceptual a priori reason not 
to consider them'' \cite{41F}. Interestingly, quantons corresponding to 
continuous spin representations exhibit many tachyonic features though 
they are not normal tachyons in the sense that they have light-like 
four-momentum \cite{41G}. Wigner's original objection against such 
``continuous spin tachyons'' is that they lead to the infinite heat capacity 
of the vacuum which can be avoided in the supersymmetric version with its 
characteristic cancellation between bosons and fermions \cite{41H}.
 
Normal tachyonic representations with spacelike four-momentum (negative mass 
squared) appear on the equal footing in the Wigner's classification. However, 
this fact does not mean that tachyons are as ubiquitous around us as bradyons 
and luxons. Let us underline that not every quanton (irreducible unitary 
representation of the Poincar\'{e} group) corresponds to localizable objects 
which can be called particles in the classical sense. Apart from the 
continuous spin representations mentioned above, we can also refer to 
the non-trivial vacuum representations of the Poincar\'{e} group with zero 
four-momentum which could correspond to pomerons \cite{41K}, queer objects
in QCD with some particle-like features (one speaks, for example, about 
pomeron exchange between protons) but nevertheless being far away from what 
is usually meant by a particle. 

Superluminality ceases to be a defining property of tachyons in quantum 
theory. When we realize this, quite a different interpretation of tachyons
emerges \cite{41AA}. In the quantum field theory to every quanton we associate 
a field $\phi$. The squared mass of the quanton is the second derivative of 
the self-interaction potential $V(\phi)$ of the field at the origin $\phi=0$. 
If the squared mass is negative, then the origin can not be the minimum of 
the potential and thus, $\phi=0$ configuration can not be a stable 
vacuum state of the theory. In other words, the system with tachyonic degree
of freedom at $\phi=0$ is unstable and the tachyonic field $\phi$ will roll
down towards the true vacuum. As the true vacuum is the minimum of the 
self-interaction potential, the squared mass is positive for the true vacuum.
Therefore, small excitations of the field $\phi$ around the true vacuum will
appear as ordinary bradyons. In fact, such a scenario is an important 
ingredient of the Standard Model and is known under the name of Higgs 
mechanism. The Higgs boson is the most famous would-be tachyon.

Interpreted in such a way, tachyons have an important revival in string theory
\cite{41AA,41BB} and in early cosmology \cite{41CC}. Even the emergence of 
time in quantum cosmology could be related to tachyons \cite{41DD}.

Summing up, tachyons do exist and play a significant role in modern quantum
theory (virtual particles, spontaneous symmetry breaking, string theory).
However, tachyons can not support the true superluminal propagation - the aim 
of their initial introduction. It can be shown that, even in a rolling state 
towards the true vacuum, localized disturbances of the tachyonic field 
never travel superluminally \cite{41CC}. ``Contrary to popular prejudice: 
{\it the tachyon is not a tachyon!}''  \cite{41CC}.

\section{Frenkel-Kontorova solitons}
Frenkel-Kontorova model \cite{41,42} describes a one-dimensional chain of 
atoms subjected to an external sinusoidal  substrate potential. The 
interactions between the nearest neighbors is assumed to be harmonic. 
Therefore, the Lagrangian of the model is
\begin{equation}
{\cal{L}}=\sum\limits_n\left\{ \frac{m}{2}\left(\frac{dx_n}{dt}\right)^2-
\frac{k}{2}(x_{n+1}-x_n-l)^2-\frac{V_0}{2}\left(1-\cos{\left(\frac{2\pi x_n}
{l}\right)}\right)\right\},
\label{eq1}
\end{equation}
where $k$ is the elastic constant of the interatomic interaction, $m$ is the 
mass of the atom, $V_0$ is the amplitude of the substrate potential and $l$
is its spatial period  which coincides with the equilibrium 
distance of the interatomic potential in our assumption. The equation of 
motion resulting from the Lagrangian (\ref{eq1}) is the following:
\begin{equation}
m\frac{d^2x_n}{dt^2}-k(x_{n+1}+x_{n-1}-2x_n)+\frac{\pi V_0}{l}\sin{\left(
\frac{2\pi x_n}{l}\right)}=0.
\label{eq2}
\end{equation} 
Let us consider the continuum limit of (\ref{eq2}) when the length $l$
characterizing the chain discreteness, is much smaller in comparison to any
relevant length scale under our interest. For this goal we introduce the
continuous variable $x$ instead of the discrete index $n$ with the 
relation $x=nl$ so that $n\pm 1$ corresponds to $x\pm l$. Besides, let us
introduce the displacements of the individual atoms from their
equilibrium positions $u_n=x_n-nl$. Note that displacements $u_n$ satisfy the 
same equation (\ref{eq2}) as the coordinates $x_n$ do. In the continuum limit, 
we can consider $u_n$ as a function of the continuous coordinate $x$ and 
expand $u_{n\pm 1}(t)\equiv u(x\pm l,t)$ in the Taylor series
$$u(x\pm l,t)\approx u(x,t)\pm\frac{\partial u}{\partial x}\,l+\frac{1}{2}\,
\frac{\partial^2 u}{\partial x^2}\,l^2.$$
Substituting this expansion into the equation (\ref{eq2}) and introducing the
dimensionless field of displacements
$$\Phi=\frac{2\pi u}{l},$$
we get the so-called sine-Gordon equation \cite{43}
\begin{equation}
\frac{1}{c^2}\,\frac{\partial^2 \Phi}{\partial t^2}-\frac{\partial^2 
\Phi}{\partial x^2}+\frac{1}{\lambda^2}\,\sin{\Phi}=0,
\label{eq3}
\end{equation}
where
\begin{equation}
c=l\sqrt{\frac{k}{m}},\;\;\;\lambda=\frac{l^2}{\pi}\sqrt{\frac{k}{2 V_0}}.
\label{eq4}
\end{equation}
For small oscillations, $\Phi\ll 1$, equation (\ref{eq3}) turns into the 
Klein-Gordon equation
\begin{equation}
\frac{1}{c^2}\,\frac{\partial^2 \Phi}{\partial t^2}-\frac{\partial^2 
\Phi}{\partial x^2}+\frac{1}{\lambda^2}\,\Phi=0
\label{eq5}
\end{equation}
describing the relativistic particle with the Compton wavelength $\lambda$.
If the external potential is switched off, $V_0\to0$, then $\lambda\to\infty$
and we get the massless phonons traveling at the speed $c$. Therefore, $c$ is 
the sound velocity for the primordial chain of atoms. In presence of the 
substrate potential, the phonons become massive and move with subsonic 
velocities (are bradyons).

We can also consider small oscillations around the point $\Phi=\pi$ which is 
the point of unstable equilibrium for the substrate potential. Writing 
$\Phi=\pi-\varphi$ and assuming $\varphi\ll 1$, we get the equation
\begin{equation}
\frac{1}{c^2}\,\frac{\partial^2 \varphi}{\partial t^2}-\frac{\partial^2 
\varphi}{\partial x^2}-\frac{1}{\lambda^2}\,\varphi=0
\label{eq6}
\end{equation}
which has the ``wrong'' sign of the mass term and describes supersonic 
phonons (tachyons). 

Interestingly, despite the supersonic behavior, (\ref{eq6}) does not allow
information to be transmitted with the velocity $v>c$. The reason is basically
the following \cite{44}: from (\ref{eq6}) we have the relation 
between the frequency $\omega$ and the wave number $k$ of the tachyonic 
excitation
$$\omega=c\sqrt{k^2-\frac{1}{\lambda^2}}.$$
If $k>1/\lambda$, the tachyonic excitations are stable. But if  $k<1/\lambda$,
$\omega$ becomes imaginary indicating the onset of instability. Nevertheless, 
any sharply localized source of perturbation (information) will have such wave
numbers in its Fourier spectrum and, therefore, any local disturbance 
inevitably will set off instability. Atoms will fall over from their
unstable $\varphi=0$ equilibrium in a domino fashion, the exponentially 
growing modes of the field $\varphi$ will quickly make the approximation 
(\ref{eq6}) inadequate and we will have to resort to the full nonlinear 
equation (\ref{eq3}) to understand what is actually happening.

So let us return to the equation (\ref{eq3}) and try to find its traveling 
wave solution $\Phi=f(x-vt)$. Substituting this traveling wave into 
(\ref{eq3}), we find that the function $f$ which determines the profile
of the wave satisfies the ordinary differential equation 
\begin{equation}
\left(1-\frac{v^2}{c^2}\right)\,\frac{d^2f}{d\xi^2}=\frac{\sin{f}}
{\lambda^2},
\label{eq7}
\end{equation}
where $\xi=x-vt$. It is easy to find the first integral of this equation in 
the form
\begin{equation}
\left(1-\frac{v^2}{c^2}\right)\,\left (\frac{df}{d\xi}\right)^2=
\frac{2}{\lambda^2}(\mu-\cos{f}),
\label{eq8}
\end{equation}
where $\mu$ is an arbitrary integration constant. Separation of variables
in (\ref{eq8}) produces in general an elliptic integral
\begin{equation}
\frac{\sigma}{L}\,(x-vt)=\int\limits_{f(0)}^f\frac{d\phi}{\sqrt{2(\mu-
\cos{\phi})}},
\label{eq9}
\end{equation}
where $\sigma=\pm 1$ and
\begin{equation}
L=\lambda\sqrt{1-\frac{v^2}{c^2}}\equiv \frac{\lambda}{\gamma}.
\label{eq10}
\end{equation}
However, if $\mu=1$, the integral (\ref{eq9}) can be calculated in terms of 
elementary functions and the result is
$$\frac{\sigma}{L}(x-vt)=\ln{\tan{\frac{f(\xi)}{4}}}-
\ln{\tan{\frac{f(0)}{4}}}.$$
Introducing $x_0$ through the relation
$$\tan{\frac{f(0)}{4}}=\exp{\left(-\frac{\sigma}{L}\,x_0\right)},$$
we get
$$\frac{\sigma}{L}(x-x_0-vt)=\ln{\tan{\frac{f(x-vt)}{4}}}.$$
As we see, $x_0$ can always be eliminated by a suitable choice of the 
coordinate origin, and so we get the following traveling wave solution
of the sine-Gordon equation
\begin{equation}
\Phi(x,t)=4\arctan{\,\exp{\left[\frac{\sigma}{L}(x-vt)\right]}}.
\label{eq11}
\end{equation} 
It is said that for $\sigma=1$ we have a kink and for $\sigma=-1$ we have 
an antikink.

But what about supersonic traveling waves? If $v>c$, then $L=i\tilde L$, 
where
\begin{equation}
\tilde L=\lambda\sqrt{\frac{v^2}{c^2}-1},
\label{eq12}
\end{equation}
and in the case of $\mu=-1$ the relation (\ref{eq9}) gives:
\begin{equation}
\Phi(x,t)=\pi-4\arctan{\,\exp{\left[-\frac{\sigma}{\tilde L}(x-vt)\right]}}.
\label{eq13}
\end{equation}
Frank and Merwe call such a tachyonic solution an anti-dislocation \cite{45}.
We will call them $T$-kink (if $\sigma=1$) and $T$-antikink (if $\sigma=-1$)
to emphasize their tachyonic nature.

In contrast to subsonic kinks, $T$-kinks are not expected to be 
stable. The reason is simple to explain \cite{46}. Ground state of the 
periodic substrate potential of the Frenkel-Kontorova model is degenerated. 
In fact we have an infinite number of different vacuum states occurring at 
$\Phi=2n\pi$, where $n$ is an integer number. To visualize this situation, 
imagine a long sheet of slate. Its depressions are just different vacuum 
states. A kink corresponds to an infinite rope which begins in one depression 
(vacuum state) and ends up in another neighboring depression (vacuum state).
Somewhere in between the rope must climb up the ridge (the maximum of the 
potential), and then fall again in different valley. The kink is stable 
because to destroy it you need to throw a rope from one valley to another
one so that it ends up completely in one vacuum state. But the rope is 
infinite and you need infinite amount of energy to perform this task.

The situation with $T$-kinks is different. $T$-kink corresponds to a rope
which lays on a potential ridge, then somewhere on the ridge it falls in the 
valley and raises again to the adjacent ridge. It is clear that such a 
configuration cannot be stable.

\section{Emergent relativity}
A remarkable fact about the Frenkel-Kontorova solitons is that they exhibit
relativistic behavior \cite{43,47}. For example, it is clear from (\ref{eq11})
that the kink is not a point-like object but an extended one and its
characteristic length is of the order of $L$. More precisely, as for 
a kink ($\sigma=1$) we have
$$\frac{1}{2\pi}\int\limits_{-\infty}^\infty \frac{\partial \Phi}{\partial x}
\,dx=\frac{1}{2\pi}\left (\Phi(\infty,t)-\Phi(-\infty,t)\right )=1,$$
and as $\Phi_x=\frac{\partial \Phi}{\partial x}$ is positive, 
symmetrically peaked around the center of the kink quantity, we can consider
$\Phi_x/2\pi$ as the spatial distribution for the kink \cite{48}. Then 
center-of-mass coordinate of the kink can be defined as \cite{48}
\begin{equation}
q=\,<x>\,=\frac{1}{2\pi}\int\limits_{-\infty}^\infty x\Phi_x\,dx,
\label{eq14}
\end{equation}
and its length as
\begin{equation}
L_q=\sqrt{<x^2>-<x>^{\,2}}.
\label{eq15}
\end{equation}
It can be easily found that
\begin{equation}
\Phi_x=\frac{2}{L}\,\frac{1}{\cosh{\frac{x-vt}{L}}},
\label{eq16}
\end{equation}
and
\begin{eqnarray} &&
<x>\,=\frac{L}{\pi}\int\limits_{-\infty}^\infty\left (y+\frac{vt}{L}\right )
\frac{dy}{\cosh{y}}=vt\,\frac{1}{\pi}\int\limits_{-\infty}^\infty
\frac{dy}{\cosh{y}}=vt, \nonumber \\ &&
<x^2>\,=\frac{L^2}{\pi}\int\limits_{-\infty}^\infty\left (y+\frac{vt}{L}
\right )^2\frac{dy}{\cosh{y}}=\frac{L^2}{\pi}\int\limits_{-\infty}^\infty
\frac{y^2}{\cosh{y}}\,dy+v^2t^2.
\label{eq17}
\end{eqnarray}
Then we obtain:
\begin{equation}
L_q=L\sqrt{\frac{1}{\pi}\int\limits_{-\infty}^\infty\frac{y^2}{\cosh{y}}\,dy}
\approx 1.57\,L,
\label{eq18}
\end{equation}
and (\ref{eq10}) shows that the kink length, $L_q$, is submitted to the 
Lorentz contraction. Interestingly, such length contraction can be observed
by naked eyes. See, for example, a strobe photography of a kink traveling
in the mechanical model of the sine-Gordon equation in \cite{43}, page 244.

Now let us consider the energy of the kink. From
$$E=\sum\limits_n\left\{ \frac{m}{2}\left(\frac{dx_n}{dt}\right)^2+
\frac{k}{2}(x_{n+1}-x_n-l)^2+\frac{V_0}{2}\left(1-\cos{\left(\frac{2\pi x_n}
{l}\right)}\right)\right\},$$
we get in the continuum limit that
\begin{equation}
E=\frac{V_0}{4}\int\limits_{-\infty}^\infty\left [\frac{\lambda^2}{c^2}
\left ( \frac{\partial \Phi}{\partial t}\right )^2+\lambda^2\left (
\frac{\partial \Phi}{\partial x}\right )^2+2(1-\cos{\Phi})\right ]\,
\frac{dx}{l},
\label{eq19}
\end{equation}
where $c$ and $\lambda$ are given by (\ref{eq4}). For the kink (\ref{eq11}) 
we have
\begin{equation}
\frac{\partial \Phi}{\partial t}=-\frac{2v}{L}\,\frac{1}{\cosh{\frac{x-vt}
{L}}},\;\;\; \frac{\partial \Phi}{\partial x}=\frac{2}{L}\,\frac{1}
{\cosh{\frac{x-vt}{L}}},
\label{eq20}
\end{equation}
and
\begin{equation}
2(1-\cos{\Phi})=\frac{16\tan^2{\frac{\Phi}{4}}}{(1+\tan^2{\frac{\Phi}{4}})^2}=
\frac{4}{\cosh^2{\frac{x-vt}{L}}}.
\label{eq21}
\end{equation}
Substituting (\ref{eq20}) and (\ref{eq21}) into (\ref{eq19}) and using the 
identity
$$1+\frac{\lambda^2}{L^2}+\frac{v^2}{c^2}\,\frac{\lambda^2}{L^2}=2\gamma^2,$$
we get
\begin{equation}
E=2\gamma^2\,\frac{L}{l}\,V_0 \int\limits_{-\infty}^\infty\frac{dy}
{\cosh^2{y}}=4\,\frac{\lambda}{l}\,V_0\gamma.
\label{eq22}
\end{equation}
Thus, we end up with the relativistic relationship between the energy and 
the mass $E=Mc^2\gamma$, where the mass of the kink is
\begin{equation}
M=4\,\frac{\lambda}{l}\,\frac{V_0}{c^2}=\frac{2}{\pi^2}\,\frac{l}{\lambda}\,
m=\frac{2m}{\pi}\sqrt{\frac{2V_0}{kl^2}}.
\label{eq23}
\end{equation}
This striking analogy between the energy of a moving single dislocation and
the energy of a particle in relativistic mechanics was first discovered by
Frenkel and Kontorova \cite{45}.

in the same way, in the case of $T$-kink we get tachyonic relations for
the $T$-kink length and energy:
\begin{equation}
L_q=L_{q0}\,\sqrt{\frac{v^2}{c^2}-1},\;\;\; E=\frac{Mc^2}{\sqrt{\frac{v^2}
{c^2}-1}},
\label{eq24}
\end{equation}
where $L_{q0}\approx 1.57\lambda$ and the mass $M$ is given again by equation
(\ref{eq23}). As we see, $T$-kinks can be considered as a mechanical model for
tachyons \cite{45}. Note that in (\ref{eq24}) it is assumed that the $T$-kink
energy is measured with respect to the potential ridge so that we have 
\begin{equation}
E=\frac{V_0}{4}\int\limits_{-\infty}^\infty\left [\frac{\lambda^2}{c^2}
\left ( \frac{\partial \Phi}{\partial t}\right )^2+\lambda^2\left (
\frac{\partial \Phi}{\partial x}\right )^2-2(1+\cos{\Phi})\right ]\,
\frac{dx}{l},
\label{eq25}
\end{equation}
instead of (\ref{eq19}).

It is clear from (\ref{eq24}) that upon loss of energy a $T$-kink will 
accelerate and become wider and wider. This paradoxical property of 
superluminal particles was deduced already by Sommerfeld just before the 
advent of special relativity \cite{39}, and it points once again towards
a transient and unstable nature of $T$-kinks. 

There is nothing particularly unexpected in emergence of the
relativistic relationships considered above because the sine-Gordon equation 
(\ref{eq3}) is Lorentz invariant excepting the fact that the light velocity 
is replaced by the sound velocity $c$. 

It is though really remarkable that this relativistic invariance is an 
emergent phenomenon. It is absent at the fundamental level (the Lagrangian 
(\ref{eq1}) is not Lorentz invariant) but appears in the long-wavelength limit.

Emergent relativity in the Frenkel-Kontorova model is approximate and holds 
only insofar as we can neglect discreteness effects. Let us return to the 
equation (\ref{eq2}) and rewrite it in following way:
$$\frac{1}{c^2}\,\frac{\partial^2\Phi}{\partial t^2}-\frac{\Phi(x+l)+
\Phi(x-l)-2\Phi(x)}{l^2}+\frac{1}{\lambda^2}\,\sin{\Phi}=0.$$
Using the Taylor expansion of the form \cite{50}
$$\Phi(x\pm l)=e^{\pm l \partial_x}\,\Phi(x),$$
where 
$$\partial_x=\frac{\partial}{\partial x},$$
we get
\begin{equation}
\Phi(x+l)+\Phi(x-l)-2\Phi(x)=2\left [\cosh{(l\partial_x)}-1\right ]\Phi(x).
\label{eq26}
\end{equation}
But
$$\cosh{x}\approx 1+\frac{x^2}{2}+\frac{x^4}{24},$$
and (\ref{eq26}) then gives
\begin{equation}
\Phi(x+l)+\Phi(x-l)-2\Phi(x)=l^2\,\left [1+\frac{l^2}{12}\,\partial_x^2
\right ]\partial_x^2\Phi(x).
\label{eq27}
\end{equation}
Therefore, we get the equation
\begin{equation}
\frac{1}{c^2}\,\frac{\partial^2\Phi}{\partial t^2}-\left(1+\frac{l^2}{12}\,
\partial_x^2\right )\,\frac{\partial^2\Phi}{\partial x^2}+\frac{1}
{\lambda^2}\,\sin{\Phi}=0.
\label{eq28}
\end{equation}
However, this equation is not convenient for considering the discreteness 
effects \cite{50,51}. For example, it contains the forth derivative of 
$\Phi$ with respect to the spatial coordinate $x$ and, hence, necessitates 
additional boundary conditions at the ends of the chain absent in the original
discrete formulation or in the zeroth-order continuum approximation. The 
remedy against this drawback is simple \cite{51}. Let us multiply (\ref{eq28}) 
by 
$$\left(1+\frac{l^2}{12}\,\partial_x^2\right )^{-1}\approx 1-\frac{l^2}{12}\,
\partial_x^2.$$
After some rearranging, we get the equation which correctly and conveniently
reproduces the first-order effects produced by the chain discreteness 
\cite{42}
\begin{equation}
 \frac{1}{c^2}\,\frac{\partial^2\Phi}{\partial t^2}-\frac{\partial^2\Phi}
{\partial x^2}+\frac{1}{\lambda^2}\,\sin{\Phi}=\frac{l^2}{12\lambda^2}\left [
\frac{\lambda^2}{c^2}\,\frac{\partial^4 \Phi}{\partial t^2 \partial x^2}+
\cos{\Phi}\,\frac{\partial^2 \Phi}{\partial x^2}-\sin{\Phi}\left(
\frac{\partial \Phi}{\partial x}\right )^2\right ].
\label{eq29}
\end{equation}
From (\ref{eq20}) it is clear that every derivative of $\Phi$ brings the 
$2\gamma/\lambda$ factor with it. Therefore, the first term in the r.h.s 
of (\ref{eq29}) is the leading one in the high energy limit and compared to 
the first term in the l.h.s, it contains an extra smallness of the order of
$$\frac{l^2\gamma^2}{3\lambda^2}=\frac{1}{3}\left(\frac{\pi^2 M\gamma}{2m}
\right)^2\sim \left(\frac{\pi Mc^2\gamma}{mc^2}\right)^2.$$
As we see, Lorentz violation remains small if the kink energy $Mc^2\gamma$
is small in comparison to the ``Plank energy'' $E_P=mc^2/\pi$.

In real life, much more significant Lorentz symmetry violation for mechanical 
kinks is caused by dissipation what brings the $\beta \Phi_t$ term in the 
equation of motion. For example, the Lorentz contraction of the kink width 
is prominent only if $\gamma\ll 1/\beta$ and it 
saturates at a value proportional to $\beta$ \cite{48} in the limit of high 
energies.

Interestingly, the breakdown of Lorentz contraction may happen even in 
Lorentz invariant theory due to quantum-field theory effects \cite{52,53,54}
and, hence, it alone does not signal a breakdown of special relativity. 
The size of an object is a classical concept and it cannot be unambiguously
extended on quantum domain. In QCD, for example, the size of the region 
which contains the information necessary to identify a hadron is determined
by fast partons and  undergoes Lorentz contraction as expected, while the low 
momentum parton cloud is universal and also determines the reasonable notion 
of the size of the hadron which however does not Lorentz contract \cite{53}. 
This leads to a very counterintuitive picture of a fast-moving nucleus being
much thinner than any of its constituent nucleons thus grossly violating our 
classical expectation that the size of a system is always larger than the
size of constituents from which the system is built \cite{54}.

\section{Supersonic solitons}
The emergent relativity in the Frenkel-Kontorova model is not universal in 
the sense that it is applied only to the excitations of the considered chain 
and does not encompass, for example, the dynamics of the substrate atoms.
This fact allows us to arrange solitons whose behavior is not restricted by 
relativistic laws. Let us consider, for example, a one-dimensional chain of 
substrate atoms with exponential interatomic interactions so that
the Lagrangian of the model is \cite{47,55} 
\begin{equation}
{\cal{L}}=\sum\limits_n\left\{ \frac{m}{2}\left(\frac{du_n}{dt}\right)^2-
\frac{k}{b}\left\{u_n-u_{n-1}+\frac{1}{b}\left[e^{-b(u_n-u_{n-1})}-1\right]
\right\}\right\}.
\label{eq30}
\end{equation}
Here again $u_n=x_n-nl$ and $m,l,k,b$ are some constants. We use the same 
notations $m,l,k$ as in the Frenkel-Kontorova model even though numerical 
values of these physical quantities may be different. The equation of motion
that follows from this Lagrangian is then
\begin{equation}
m\frac{d^2u_n}{dt^2}+\frac{k}{b}\left[ e^{-b(u_{n+1}-u_n)}-
e^{-b(u_n-u_{n-1})}\right ]=0.
\label{eq31}
\end{equation}
Note that the case of small $b$ corresponds to the harmonic interatomic 
interactions.

To find a solitonic solution of (\ref{eq31}) we can proceed as follows 
\cite{56}. Let us introduce dimensionless variables $w_n$ and $\tau$ through
relations 
\begin{equation}
\tau=\sqrt{\frac{k}{m}}\,t,
\label{eq32}
\end{equation}
and
\begin{equation}
1+\dot{w}_n=e^{-b(u_n-u_{n-1})},\;\;\;\;w_n-w_{n+1}=b\,\dot{u}_n.
\label{eq33}
\end{equation}
Here the dot indicates differentiation with respect to $\tau$ so that
$$\dot{u}_n=\sqrt{\frac{m}{k}}\,\frac{du_n}{dt}.$$

By differentiation of the second equation in (\ref{eq33}) with respect to 
time $t$ and with help of the first one, it is easy to check that $u_n$ 
satisfies indeed the original equation (\ref{eq31}). On the other hand, we 
have an equation for $w_n$ following from (\ref{eq33}):
\begin{equation}
\frac{\ddot{w}_n}{1+\dot{w}_n}=w_{n+1}+w_{n-1}-2w_n,
\label{eq34}
\end{equation}
and we can find its solitonic solution by a B\"{a}cklund transformation 
\cite{56}.

In differential geometry, the B\"{a}cklund transformation enables the 
construction of a new pseudospherical surface (a surface with a constant and 
negative Gaussian curvature) from a given pseudospherical surface. 
Technically the B\"{a}cklund transformation is a pair of first order partial 
differential equations which relate two different solutions of the second 
order partial differential equations. This transformation has important 
applications in soliton theory \cite{57}. 

Toda and Wadati extended the idea to a differential-difference equations 
and obtained a discrete analog of B\"{a}cklund transformation for
the exponential lattice \cite{58}. For the equation (\ref{eq34}) the 
B\"{a}cklund transformation was found in \cite{56} and  it has the form
\begin{eqnarray} &&
1+\dot{w}_n=(\lambda+w_n^\prime-w_n)(\lambda+w_n-w_{n+1}^\prime),
\nonumber \\ &&
1+\dot{w}_n^\prime=(\lambda+w_n^\prime-w_n)(\lambda+w_{n-1}-w_n^\prime),
\label{eq35}
\end{eqnarray}
where $\lambda$ is an arbitrary constant. If $w_n(t)$ and $w_n^\prime(t)$
are any two functions related by (\ref{eq35}), they both are solutions of the 
equation (\ref{eq34}). For example, we have from (\ref{eq35})
\begin{equation}
\frac{\ddot{w}_n^\prime}{1+\dot{w}_n^\prime}=\frac{d}{d\tau}\ln{(1+\dot{w}_n
^\prime)}=\frac{\dot{w}_n^\prime-\dot{w}_n}{\lambda+w_n^\prime-w_n}+
\frac{\dot{w}_{n-1}-\dot{w}_n^\prime}{\lambda+w_{n-1}-w_n^\prime}.
\label{eq36}
\end{equation}
However, again from (\ref{eq35}),
\begin{eqnarray} &&
\dot{w}_n^\prime-\dot{w}_n=(\lambda+w_n^\prime-w_n)(w_{n-1}-w_n^\prime-w_n+
w_{n+1}^\prime), \nonumber \\ &&
\dot{w}_{n-1}-\dot{w}_n^\prime=(\lambda+w_{n-1}-w_n^\prime)(w_{n-1}^\prime-
w_{n-1}-w_n^\prime+w_n).
\label{eq37}
\end{eqnarray}
Substituting (\ref{eq37}) into (\ref{eq36}), we see that $w_n^\prime$ is 
indeed a solution of (\ref{eq34}):
$$\frac{\ddot{w}_n^\prime}{1+\dot{w}_n^\prime}=w_{n+1}^\prime+w_{n-1}^\prime
-2w_n^\prime.$$
Let $w_n^\prime=0$ be a trivial solution of (\ref{eq34}). Then (\ref{eq35}) 
takes the form
\begin{equation}
1+\dot{w}_n=\lambda^2-w_n^2,\;\;\;\;1=(\lambda-w_n)(\lambda+w_{n-1}).
\label{eq38}
\end{equation}
We further assume that $\lambda^2\ge 1$ so that we can write $\lambda=\pm
\cosh{\phi}$ for some $\phi$. The first equation in (\ref{eq38}) can easily
be integrated then:
\begin{equation}
w_n=\sinh{\phi}\,\tanh{[\tau\,\sinh{\phi}+\alpha_n]},
\label{eq39}
\end{equation}
where $\alpha_n$ is the integration constant which is the only quantity  
in (\ref{eq39}) that can depend on $n$. Using (\ref{eq39}) and the identity
$\tanh{x}-\tanh{y}=\tanh{(x-y)}\,[1-\tanh{x}\,\tanh{y}]$, we obtain that the 
second equation
of (\ref{eq38}) can be rewritten in the form
$$\sinh^2{\phi}=$$ 
$$\sinh^2{\phi}\,\tanh{x}\,\tanh{y}-\sinh{\phi}\,\cosh{\phi}\,
\tanh{(\alpha_n-\alpha_{n-1})}\,[1-\tanh{x}\,\tanh{y}],$$
where $x=\tau\,\sinh{\phi}+\alpha_n$, $y=\tau\,\sinh{\phi}+\alpha_{n-1}$ and
for definiteness we have taken $\lambda=-\cosh{\phi}$. This identity must be 
valid for any $\tau$. It is possible only if 
$$\cosh{\phi}\,\tanh{(\alpha_n-\alpha_{n-1})}=-\sinh{\phi}.$$
Consequently, $\alpha_n-\alpha_{n-1}=-\phi$ what implies 
$\alpha_n=-n\phi+\alpha_0.$
Finally, we get the following nontrivial solution of (\ref{eq34})
\begin{equation}
w_n=\sinh{\phi}\,\tanh{[\tau\,\sinh{\phi}-n\phi+\alpha_0]}.
\label{eq40}
\end{equation}
In the continuum limit with $x=nl$ we have 
\begin{equation}
w(x,t)=\sinh{\frac{l}{L}}\,\tanh{\frac{vt-x+x_0}{L}},
\label{eq41}
\end{equation}
where
\begin{equation}
L=\frac{l}{\phi},\;\;\;x_0=\alpha_0L,\;\;\;v=c\,\frac{L}{l}\,
\sinh{\frac{l}{L}},
\label{eq42}
\end{equation}
and $c=\sqrt{k/m}\,l$ is the sound velocity for the harmonic chain (in the 
limit $b\to 0$). Note that the Toda soliton (\ref{eq41}) is supersonic,
$v>c$, since $\sinh{x}>x$ for any $x>0$.

It is clear from (\ref{eq41}) that the soliton width is of the order of $L$.
The continuum approximation assumes $L\gg l$, then the soliton is only 
slightly supersonic 
$$\left (\frac{v}{c}\right )^2=\frac{\sinh^2{\phi}}{\phi^2}=\frac{1}{2\phi^2}
\,(\cosh{2\phi}-1)\approx 1+\frac{1}{3}\frac{l^2}{L^2},$$
and its width depends on the velocity as follows
\begin{equation}
L=\frac{l/\sqrt{3}}{\sqrt{\frac{v^2}{c^2}-1}}.
\label{eq43}
\end{equation}
From (\ref{eq33}) we get in the continuum limit when $w\ll 1$,
\begin{equation}
\frac{du_n}{dt}\approx -\frac{c}{b}\,\frac{\partial w}{\partial x},\;\;
u_n-u_{n-1}\approx -\frac{1}{b}\,\sqrt{\frac{m}{k}}\,\frac{\partial w}
{\partial t}=\frac{v}{c}\,\frac{l}{b}\,\frac{\partial w}{\partial x}.
\label{eq44}
\end{equation}
Therefore, in the harmonic approximation for the 
potential energy, the energy of the soliton is
$$E\approx \frac{m}{2b^2}\,(v^2+c^2)\int\limits_{-\infty}^\infty \left(
\frac{\partial w}{\partial x}\right )^2\frac{dx}{l}=$$
\begin{equation}
\frac{m}{2b^2}\,(v^2+c^2)
\,\frac{l}{L^3}\int\limits_{-\infty}^\infty \frac{dy}{\cosh^4{y}}=
\frac{2l}{3b^2L^3}\,m\,(v^2+c^2).
\label{eq45}
\end{equation}
As we see from (\ref{eq43}) and (\ref{eq45}), when the velocity of the Toda 
soliton approaches the sound velocity, its energy turns to zero and its
width turns to infinity. Such a behavior is opposite to that of a tachyon but
the result is the same: the Toda soliton cannot cross the sound barrier and 
become subsonic. But there are other types of solitons which can: 
generalized Frenkel-Kontorova model with the special kind of anharmonicity
is a specific example \cite{59}. The Lagrangian of the model is
\cite{42,49,59}  
$${\cal{L}}=\sum\limits_n\left\{ \frac{m}{2}\left(\frac{du_n}{dt}\right)^2-
\right .$$
\begin{equation}
\left .
\frac{k}{2}(u_{n+1}-u_n)^2\left [1+\frac{\chi}{l^2}(u_{n+1}-u_n)^2\right ]
-\frac{V_0}{2}\left(1-\cos{\left(\frac{2\pi u_n}{l}\right)}\right)\right\},
\label{eq46}
\end{equation}
where $\chi$ is a dimensionless anharmonicity parameter. Correspondingly,
the equations of motion are
$$m\frac{d^2u_n}{dt^2}-k(u_{n+1}+u_{n-1}-2u_n)-\frac{2k\chi}{l^2}\left [
(u_{n+1}-u_n)^3-(u_n-u_{n-1})^3\right ]+$$
\begin{equation}
\frac{\pi V_0}{l}\sin{\left(
\frac{2\pi x_n}{l}\right)}=0.
\label{eq47}
\end{equation} 
To get the continuum limit, let us expand these equations up to terms of the 
order of $l^5$
(assuming that $u_n$ and its spatial derivatives are of the order of $l$):
\begin{eqnarray}  &&
u_{n+1}+u_{n-1}-2u_n\approx l^2\,\frac{\partial^2 u_n}{\partial x^2}+
\frac{l^4}{12}\,\frac{\partial^4 u_n}{\partial x^4}, \nonumber \\ &&
\frac{1}{l^2}\left [(u_{n+1}-u_n)^3-(u_n-u_{n-1})^3\right ] \approx
3l^2\,\left(\frac{\partial u_n}{\partial x}\right )^2\,\frac{\partial^2 u_n}
{\partial x^2}.
\label{eq48}
\end{eqnarray}
Therefore, the continuum limit of (\ref{eq47}) is given by
\begin{equation}
\frac{1}{c^2}\,\frac{\partial^2 \Phi}{\partial t^2}-\frac{\partial^2 
\Phi}{\partial x^2}-\frac{l^2}{12}\,\frac{\partial^4 \Phi}{\partial x^4}-
\frac{3\chi l^2}{2\pi^2}\,\left(\frac{\partial\Phi}{\partial x}\right)^2\,
\frac{\partial^2 \Phi}{\partial x^2}+\frac{1}{\lambda^2}\,\sin{\Phi}=0,
\label{eq49}
\end{equation} 
where, as before, $\Phi$ is given by $$\Phi=\frac{2\pi u}{l}.$$

For the special value of the anharmonicity parameter $\chi$, equation 
(\ref{eq49}) has solitonic solutions of the same functional form (\ref{eq11}) 
as in the harmonic case. But the velocity dependence of $L$ is not given by
(\ref{eq10}) and has a more complicated form. Indeed, (\ref{eq20}) and
(\ref{eq21}) indicate that we have the following relations 
$$\frac{\partial^2 \Phi}{\partial t^2}=v^2\,\frac{\partial^2 \Phi}
{\partial x^2},\;\;\;\frac{\partial^2 \Phi}{\partial x^2}=\frac{\sin{\Phi}}
{L^2},\;\;\;\left (\frac{\partial \Phi}{\partial x}\right)^2=\frac{2(1-
\cos{\Phi})}{L^2},$$
from which we get:
\begin{eqnarray} &&
\frac{\partial^4 \Phi}{\partial x^4}=-\frac{2\sin{\Phi}}{L^4}+
\frac{3\sin{2\Phi}}{2L^4}, \nonumber \\ &&
\left(\frac{\partial \Phi}{\partial x}\right )^2\,\frac{\partial^2 \Phi}
{\partial x^2}=\frac{2\sin{\Phi}}{L^4}-\frac{\sin{2\Phi}}{L^4}.
\label{eq50}
\end{eqnarray}
It follows from (\ref{eq50}) that if
\begin{equation}
\chi=\frac{\pi^2}{12},
\label{eq51}
\end{equation}
then
$$\frac{l^2}{12}\,\frac{\partial^4 \Phi}{\partial x^4}+\frac{3\chi l^2}
{2\pi^2}\,\left(\frac{\partial \Phi}{\partial x}\right )^2\,\frac{\partial^2 
\Phi}{\partial x^2}=\frac{l^2}{12}\,\frac{\sin{\Phi}}{L^4},$$
and (\ref{eq11}) will be a solution of (\ref{eq49}) if
$$\frac{v^2/c^2-1}{L^2}-\frac{l^2}{12L^4}+\frac{1}{\lambda^2}=0,$$
or 
\begin{equation}
L^4-\left(1-\frac{v^2}{c^2}\right )\,\lambda^2\,L^2- \frac{l^2\lambda^2}{12}
=0.
\label{eq52}
\end{equation}
The positive solution of (\ref{eq52}) is
\begin{equation}
L^2=\frac{\lambda^2}{2}\left [ 1-\frac{v^2}{c^2}+\sqrt{\left( 1-\frac{v^2}
{c^2}\right )^2+\frac{1}{3}\,\left(\frac{l}{\lambda}\right )^2}\right ].
\label{eq53}
\end{equation}
Note that (\ref{eq53}) remains finite and nonzero at $v=c$. Therefore, there 
is no sonic barrier for this type of solitons (Kosevich-Kovalev solitons). 
In a sense, Kosevich-Kovalev soliton interpolates  between the subsonic 
Frenkel-Kontorova solitons and supersonic Toda solitons \cite{42,49}. Indeed,
for $v\gg c$ we get from (\ref{eq53})
$$L=\frac{l/2\sqrt{3}}{\sqrt{\frac{v^2}{c^2}-1}},$$
which is half the width of the Toda soliton. While for  $l/\lambda\ll 1$ and
$v\ll c$ we have the same width as for the Frenkel-Kontorova solitons:
$$L=\lambda\,\sqrt{1-\frac{v^2}{c^2}}.$$
In contrast to Frenkel-Kontorova and Toda solitons, Kosevich-Kovalev solitons
can move with any velocity from zero to infinity.

\section{Elvisebrions}
We believe ``that theory acquires  authority by confronting and conforming 
to experiment, not the other way around'' \cite{60}. To quote Richard 
Feynman, ``it does not make any difference how beautiful your guess is. It 
does not make any difference how smart you are, who made the guess, or what 
his name is -- if it disagrees with experiment it's wrong'' \cite{60A}. 
Special relativity is an idea that  was scrutinized
experimentally many times and always found to be conforming to experiment. 
However, ``history of physics shows that with the unique exception of current 
laws and theories, all previous hypotheses have been surpassed by the new 
order introduced and that, subsequently, they have been proved wrong or 
limited in some way or another'' \cite{40D}. Why should special relativity
be an exception?

Frenkel-Kontorova model is a simple mechanical example which hints toward
a possibility that special relativity might be actually an emergent 
phenomenon: valid only when things are inspected at relevant scales but 
disappears at finer scales. In the realistic Frenkel-Kontorova model,
relativity disappears both in the short wave-length limit (due to discreteness
effects) and in the very long-wave-length limit (due to finiteness of the
chain). Interestingly, superfluid $^3He$-$A$ provides another and even more
interesting and realistic example where the relativistic quantum field theory
emerges as the effective theory in the low energy corner but, at the same time,
the limiting behaviour for high and ultralow energies contradicts special
relativity theory \cite{61}.

There are several reasons for why we should take the idea of emergent 
relativity seriously. The Frenkel-Kontorova model is just only one example of 
a relativistic behavior which emerges in purely classical-mechanical systems 
\cite{47}. In quantum world such examples proliferate. All ingredients of
the Standard Model, such as chiral fermions, Lorentz symmetry, gauge 
invariance, chiral anomaly, have their counterparts as emergent phenomena in 
condensed matter physics \cite{62}. Last but not least, it seems that 
emergence of hierarchies of laws is the basic principle of Nature's 
functionality \cite{60,63}. All our experience in physics confirms this  
basic principle, especially in condensed matter physics ``where theoretical 
ideas are forced to immediate and brutal confrontation with experiment by 
virtue of the latter's low cost'' \cite{60}.

However, if the special relativity is indeed an emergent phenomenon then there
may exist a ``substrate'' whose excitations do not belong to the relativistic
world and, therefore, can move superluminally. It is clear that such type of
superluminal particle-like excitations of the substrate, analogs of Toda 
or Kosevich-Kovalev solitons, are conceptually different from tachyons and
deserve their own name. We name them ``elvisebrions'' ({\mxedr elvisebri}
- elvisebri in Georgian means ``swift as a lightning flash''. Hopefully,
admirers of the Elvis Presley music will also appreciate the name). 

Giving a name to something already brings it into a kind of 
existence, ``it is made at least virtually real'' \cite{64}. However, is this 
existence more substantial than that of unicorns? Only experiment can tell. 
We believe that it is at least worthwhile to continue the search of 
superluminal  particles. Probability of success is hard to estimate, but we 
can refer to Alvarez principle to justify such a research (the argument is 
taken from \cite{65} where it was applied to the search of tachyons). The 
Alvarez principle relates the merit of an experiment, $\mu$, to the 
probability of its success, $P$, and to the significance of the result, 
$\sigma$, in the following way: $\mu=\sigma\cdot P$. We suspect that most 
physicists, in their sound mind, will insist that for 
elvisebrions $P=0$. Nevertheless, they probably will agree that in the case 
of positive result, $\sigma=\infty$, and in Calculus $0\cdot \infty$ is 
indeterminate. In the case of indeterminate $\mu$, everything rests on 
``the gumption of the experimenters'' \cite{65}.   

To be a bit more specific what we have in mind when speaking about 
elvisebrions, let us consider the dynamics of interacting Frenkel-Kontorova
and Kosevich-Kovalev chains given by the Lagrangian
\begin{equation}
{\cal L}={\cal L}_1+{\cal L}_2+{\cal L}_{int},
\label{eq54}
\end{equation}
where ${\cal L}_1$ is given by (\ref{eq1}) with changes $m\to m_1$, $k\to
k_1$, $V_0\to V_1$ and ${\cal L}_2$ is given be (\ref{eq46}) with changes 
$m\to m_2$, $k\to k_2$, $V_0\to V_2$ and $u_n\to v_n=y_n-nl$, while
\begin{equation}
{\cal L}_{int}=-\frac{V_0}{2}\left(1-\cos{\left(\frac{2\pi (x_n-y_n)}{l}
\right)}\right ). 
\label{eq55}
\end{equation}
In the long-wavelength limit, and assuming that
\begin{equation}
\frac{k_1}{k_2}=\frac{m_1}{m_2}=\frac{V_1}{V_2}=\frac{V_0}{V_1},
\label{eq56}
\end{equation}
we get the following  system of coupled equations
\begin{eqnarray} &&
\frac{1}{c^2}\,\frac{\partial^2 \Phi}{\partial t^2}-\frac{\partial^2 
\Phi}{\partial x^2}+\frac{1}{\lambda^2}\,\sin{\Phi}=\frac{1}{\lambda^2}\,
\frac{m_1}{m_2}\,\sin{(\Psi-\Phi)},\nonumber \\ &&
\frac{1}{c^2}\,\frac{\partial^2 \Psi}{\partial t^2}-\frac{\partial^2 
\Psi}{\partial x^2}-\frac{l^2}{12}\,\frac{\partial^4 \Psi}{\partial x^4}-
\frac{l^2}{8}\,\left(\frac{\partial\Psi}{\partial x}\right)^2\,
\frac{\partial^2 \Psi}{\partial x^2}+\frac{1}{\lambda^2}\,\sin{\Psi}=
\nonumber \\ &&
\frac{1}{\lambda^2}\,\left(\frac{m_1}{m_2}\right)^2\sin{(\Phi-\Psi)}, 
\label{eq57}
\end{eqnarray}
where $\Psi$ is the dimensionless field of $y$-displacements defined 
analogously to $\Phi$, and
\begin{eqnarray} &&
c=l\sqrt{\frac{k_1}{m_1}},\;\;\lambda=\frac{l^2}{\pi}\sqrt{\frac{k_1}
{2 V_1}}.
\label{eq58}
\end{eqnarray}
Now suppose $m_1\ll m_2$. Then, in the zeroth approximation the dynamics 
of chains decouple and $\Phi$-excitations live in a relativistic sine-Gordon
world unaware of the existence of a hidden elvisebrion  $\Psi$-sector which 
is not Lorentz invariant. In the first approximation however, the decoupling 
is not complete and while we do still have the Lorentz invariant $\Psi=0$ 
sector and no possibility for $\Phi$-inhabitants of this sector to excite 
(in this approximation) $\Psi$ degrees of freedom, the opposit is not 
correct: $\Psi$-excitations are coupled (albeit weakly) to the $\Phi$-sector 
and therefore their presence can be detected by $\Phi$-inhabitants.

Note that the system (\ref{eq57}) still has a supersonic solution
\begin{eqnarray} &&
\Psi(x,t)=4\arctan{\,\exp{\left[\frac{\sigma}{L}(x-vt)\right]}}, 
\nonumber \\ &&
\Phi(x,t)=\pi-4\arctan{\,\exp{\left[-\frac{\sigma}{L}(x-vt)\right]}}=
\nonumber \\ &&
-\pi+4\arctan{\,\exp{\left[\frac{\sigma}{L}(x-vt)\right]}},
\label{eq59}
\end{eqnarray}
provided that
$$L^2=\lambda^2\left(\frac{v^2}{c^2}-1\right)=
\frac{\lambda^2}{2}\left [ 1-\frac{v^2}{c^2}+\sqrt{\left( 1-\frac{v^2}
{c^2}\right )^2+\frac{1}{3}\,\left(\frac{l}{\lambda}\right )^2}\right ]$$
which for the velocity $v$ gives
\begin{equation}
\frac{v^2}{c^2}=1+\frac{1}{\sqrt{24}}\,\frac{l}{\lambda}.
\label{eq60}
\end{equation}
It is also interesting to consider an opposite limit $m_1\gg m_2$ of 
(\ref{eq57}). To simplify the treatment, we neglect the anharmonicity this 
time while relaxing the (\ref{eq56}) condition to the following one 
(introducing a dimensionless parameter $\epsilon$)
\begin{equation}
\epsilon^2\,\frac{k_1}{k_2}=\frac{m_1}{m_2}=\frac{V_1}{V_2}=\frac{V_0}{V_1}.
\label{eq61}
\end{equation}
Then, in the limit $m_1\gg m_2$, we obtain a system of coupled sine-Gordon 
equations
\begin{eqnarray} &&
\frac{1}{c^2}\,\frac{\partial^2 \Phi}{\partial t^2}-
\frac{\partial^2 \Phi}{\partial x^2}=\frac{1}{\lambda^2}\,\frac{m_1}
{m_2}\,\sin{(\Psi-\Phi)},\nonumber \\ &&
\frac{1}{(\epsilon c)^2}\,\frac{\partial^2 \Psi}{\partial t^2}-
\frac{\partial^2 \Psi}{\partial x^2}=\frac{1}{\lambda^2}\left(\frac{m_1}
{\epsilon m_2}\right )^2\sin{(\Phi-\Psi)}.
\label{eq62}
\end{eqnarray}
Note the physical meaning of the parameter $\epsilon$:
\begin{equation}
\epsilon=\frac{c_2}{c_1},
\label{eq63}
\end{equation}
where $c_1$ and $c_2$ are sound velocities associated to the individual 
Frenkel-Kontorova chains.

It will be convenient to introduce dimensionless variables instead of $t$
and $x$ through
\begin{equation}
\tilde{x}=\frac{x}{\lambda}\,\frac{m_1}{m_2},\;\;\;
\tilde{t}=\frac{ct}{\lambda}\,\frac{m_1}{m_2}.
\label{eq64}
\end{equation}
Then we get from (\ref{eq62})
\begin{eqnarray} &&
\frac{\partial^2 \Phi}{\partial \tilde{t}^2}-
\frac{\partial^2 \Phi}{\partial \tilde{x}^2}=-\delta^2\sin{(\Phi-\Psi)},
\nonumber \\ &&
\frac{\partial^2 \Psi}{\partial \tilde{t}^2}-\epsilon^2\,
\frac{\partial^2 \Psi}{\partial \tilde{x}^2}=\sin{(\Phi-\Psi)},
\label{eq65}
\end{eqnarray}
where
\begin{equation}
\delta^2=\frac{m_2}{m_1}.
\label{eq66}
\end{equation}
The system (\ref{eq65}) is a generalization of the Frenkel-Kontorova model
of crystal dislocations \cite{66}. Indeed, in the limit $\delta\to 0$ 
($m_1\gg m_2$) and for $\Phi=0$, we obtain for $\Psi$ the sine-Gordon equation 
for the long-wavelength description of the displacements of the particles of 
mass $m_2$ while treating the much heavier particles of mass $m_1$ as 
motionless. Interestingly, the system (\ref{eq65}) with $\epsilon=1$ was 
suggested to describe soliton excitations in deoxyribonucleic acid (DNA) 
double helices \cite{66A}. In the general case $\epsilon\ne 1,\;\delta\ne 0$, 
solutions of (\ref{eq65}) were considered in \cite{66,67,68}. When 
$\epsilon\ne 1$ (that is when $c_1$ and $c_2$ sound velocities are different), 
the coupled system (\ref{eq65}) is not Lorentz invariant. As a result, 
traveling wave solutions do appear with supersonic velocities, as we now 
demonstrate, following \cite{66}.

Let us seek the solution of (\ref{eq65}) in the form
\begin{equation}
\Phi(\tilde{x},\tilde{t})=A\,p(\xi),\;\;\;\;
\Psi(\tilde{x},\tilde{t})=B\,p(\xi),
\label{eq67}
\end{equation}
where $\xi=\tilde{x}-\nu\tilde{t}$, and $\nu$, $A$, $B$ are some constants 
such that 
\begin{equation}
A-B=1.
\label{eq68}
\end{equation}
Substituting (\ref{eq67}) into (\ref{eq65}), we get
\begin{eqnarray} &&
A(\nu^2-1)\,\frac{d^2p}{d\xi^2}=-\delta^2\,\sin{p}, \nonumber \\ &&
B(\nu^2-\epsilon^2)\,\frac{d^2p}{d\xi^2}=\sin{p}.
\label{eq69}
\end{eqnarray}
Therefore,
\begin{equation}
\frac{A(\nu^2-1)}{B(\nu^2-\epsilon^2)}=-\delta^2,
\label{eq70}
\end{equation}
and this relation, together with (\ref{eq68}), determines $A$ and $B$ 
constants as 
\begin{equation}
A=\frac{\delta^2 (\nu^2-\epsilon^2)}{\delta^2 (\nu^2-\epsilon^2)+\nu^2-1},
\;\;\;\;
B=\frac{1-\nu^2}{\delta^2 (\nu^2-\epsilon^2)+\nu^2-1},
\label{eq71}
\end{equation}
while for $p(\xi)$ we get the equation
\begin{equation}
\frac{d^2p}{d\xi^2}+\mu\,\sin{p}=0,
\label{eq72}
\end{equation}
where
\begin{equation}
\mu=\frac{\delta^2 (\nu^2-\epsilon^2)+\nu^2-1}{(\nu^2-\epsilon^2)(\nu^2-1)}=
\frac{\delta^2}{\nu^2-1}+\frac{1}{\nu^2-\epsilon^2}.
\label{eq73}
\end{equation}
We will assume $\epsilon<1$ and first consider the case $\mu>0$ which is only 
possible if
\begin{equation}
\epsilon^2<\nu^2<\frac{1+\delta^2\,\epsilon^2}{1+\delta^2},\;\;\;
\mathrm{or}\;\;\; 1<\nu^2<\infty.
\label{eq74}
\end{equation}
Assuming 
$$\frac{dp}{d\xi}(\xi=0)=0,\;\;\;\mathrm{and}\;\;\;p(\xi=0)=p_0,$$
we have from (\ref{eq72}) the following first integral
$$\frac{1}{2}\left (\frac{dp}{d\xi}\right)^2=\mu(\cos{p}-\cos{p_0})=
2\mu\left (\sin^2{\frac{p_0}{2}}-\sin^2{\frac{p}{2}}\right).$$
Therefore,
\begin{equation}
\int\limits_{p_0}^p\frac{d(p/2)}{\sqrt{\sin^2{\frac{p_0}{2}}-
\sin^2{\frac{p}{2}}}}=\sigma\sqrt{\mu}\,\xi,
\label{eq75}
\end{equation}
with $\sigma=\pm 1$. Let us make the following substitution in the integral:
$$\sin{\frac{p}{2}}=k\,\sin{\phi},\;\;\;k=\sin{\frac{p_0}{2}}.$$
Then we get
$$\int\limits_{\pi/2}^\phi \frac{d\phi}
{\sqrt{1-k^2\,\sin^2{\phi}}}=\sigma\sqrt{\mu}\,\xi,$$
and, therefore,
\begin{equation}
\mathrm{sn}^{-1}\left(\frac{1}{k}\,\sin{\frac{p}{2}},k\right)-K(k)=
\sigma\sqrt{\mu}\,(\tilde{x}-\nu\tilde{t}),
\label{eq76}
\end{equation}
where $\mathrm{sn}(u,k)$ is one of the Jacobian elliptic functions (for
elementary theory of elliptic functions see, for example, \cite{68A})
and $K(k)$ is the complete elliptic integral of the first kind. It is clear 
from (\ref{eq76}) that $K(k)$ can be absorbed by the redifinition of the 
spatial origin and we ignore it in the following.

Therefore, the case $\mu>0$ corresponds to the fast traveling waves \cite{66}
which can propagate with velocity $v=\nu c\equiv \nu c_1$ in the range
\begin{equation}
c_2^2<v^2<\frac{m_1c_1^2+m_2c_2^2}{m_1+m_2},\;\;\;\mathrm{or}\;\;\;
c_1^2<v^2<\infty,
\label{eq77}
\end{equation} 
and have the form (we have switched back from dimensionless quantities)
\begin{eqnarray} &&
\Phi(x,t)=\frac{2\,\delta^2\left(\frac{v^2}{c^2}-\epsilon^2\right)}
{\delta^2\left(\frac{v^2}{c^2}-\epsilon^2\right)+\frac{v^2}{c^2}-1}\,
\arcsin{\left[k\,\mathrm{sn}\left(\frac{\sigma}{\tilde L}(x-vt),\,k\right)
\right]}, \nonumber\\ &&
\Psi(x,t)=\frac{2\left(1-\frac{v^2}{c^2}\right)}
{\delta^2\left(\frac{v^2}{c^2}-\epsilon^2\right)+\frac{v^2}{c^2}-1}\,
\arcsin{\left[k\,\mathrm{sn}\left(\frac{\sigma}{\tilde L}(x-vt),\,k\right)
\right]},
\label{eq78}
\end{eqnarray} 
where
\begin{equation}
\tilde L=\frac{\lambda\delta^2}{\sqrt{\mu}}=\lambda\delta^2\sqrt{\frac
{\left (\frac{v^2}{c^2}-\epsilon^2\right )\left (\frac{v^2}{c^2}-1\right )}
{\delta^2 \left (\frac{v^2}{c^2}-\epsilon^2\right)+\frac{v^2}{c^2}-1}}.
\label{eq79}
\end{equation} 
Now consider the case $\mu<0$, that is (remember that we have assumed
$\epsilon<1$)
\begin{equation}
0\le \nu^2<\epsilon^2,\;\;\;\mathrm{or}\;\;\; \frac{1+\delta^2\,\epsilon^2}
{1+\delta^2}<\epsilon^2<1.
\label{eq80}
\end{equation}
If we take $p=\tilde p -\pi$, then the equation for $\tilde p$ will be
\begin{equation}
\frac{d^2\tilde p}{d\xi^2}+(-\mu)\,\sin{\tilde p}=0.
\label{eq81}
\end{equation}
That is for $\tilde p$ we have the previously discussed case $\tilde \mu=
-\mu>0$ and, therefore, we already know the solution of (\ref{eq81}). Using
the relation $\mathrm{dn}^2(u,k)=1-k^2\mathrm{sn}^2(u,k)$ among Jacobi 
elliptic functions, we finally get a slow traveling wave solution which
can propagate with velocities in the range
\begin{equation}
0\le v^2<c_2^2,\;\;\;\mathrm{or}\;\;\;
\frac{m_1c_1^2+m_2c_2^2}{m_1+m_2}<v^2<c_1^2,
\label{eq82}
\end{equation} 
and have the form
\begin{eqnarray} &&
\Phi(x,t)=\frac{2\,\delta^2\left(\frac{v^2}{c^2}-\epsilon^2\right)}
{\delta^2\left(\frac{v^2}{c^2}-\epsilon^2\right)+\frac{v^2}{c^2}-1}\,
\arcsin{\left[\mathrm{dn}\left(\frac{\sigma}{L}(x-vt),\,k\right)
\right]}, \nonumber\\ &&
\Psi(x,t)=\frac{2\left(1-\frac{v^2}{c^2}\right)}
{\delta^2\left(\frac{v^2}{c^2}-\epsilon^2\right)+\frac{v^2}{c^2}-1}\,
\arcsin{\left[\mathrm{dn}\left(\frac{\sigma}{L}(x-vt),\,k\right)
\right]},
\label{eq83}
\end{eqnarray} 
where
\begin{equation}
L=\frac{\lambda\delta^2}{\sqrt{-\mu}}=\lambda\delta^2\sqrt{\frac
{\left (\epsilon^2-\frac{v^2}{c^2}\right )\left (\frac{v^2}{c^2}-1\right )}
{\delta^2 \left (\frac{v^2}{c^2}-\epsilon^2\right)+\frac{v^2}{c^2}-1}}.
\label{eq84}
\end{equation} 
Note that the slow traveling waves may move with velocities which can exceed
$c_2$, the limiting velocity for the second Frenkel-Kontorova chain. 
This waves are a generalization of the Frenkel-Kontorova solitons 
(\ref{eq11}), while the fast  traveling waves, (\ref{eq78}), are 
a generalization of the tachyonic anti-dislocations (\ref{eq13}). 
Indeed, let us consider the limit
\begin{equation}
k\to 1,\;\;\;\epsilon\to 1,\;\;\;\delta\to 0,\;\;\;
\lambda\,\delta\to \lambda_0.
\label{eq85}
\end{equation}
Using
$$\arcsin{x}=2\,\arctan{\frac{x}{1+\sqrt{1-x^2}}},$$
and the relation  $\mathrm{dn}(u,1)=\mathrm{sech}\,u$ \cite{68A}, we get 
for the slow traveling waves, in the limit (\ref{eq85}), $\Phi=0$ and
$$\Psi=-4\arctan{\,\exp{\left[-\frac{\sigma}{L}(x-vt)\right]}}=
-2\pi+\arctan{\,\exp{\left[\frac{\sigma}{L}(x-vt)\right]}},$$
where, according to (\ref{eq84}), $L$ takes the form (\ref{eq10}), with
$\lambda_0$ instead of $\lambda$, in this limit.

As for the fast traveling waves, we can use  the relation $\mathrm{sn}(u,1)=
\tanh{u}$ \cite{68A} to get from (\ref{eq78}), in the limit (\ref{eq85}), 
$$\sin{\frac{\Psi}{2}}=\tanh{\left (-\frac{\sigma}{\tilde L}(x-vt)\right )},$$
and therefore
$$\exp{\left (-\frac{\sigma}{\tilde L}(x-vt)\right )}=
\sqrt{\frac{1-\sin{\frac{\Psi}{2}}}{1-\sin{\frac{\Psi}{2}}}}=
\tan{\left (\frac{\pi}{4}-\frac{\Psi}{4}\right )},$$
from which it follows that $\Psi$ is an anti-dislocation (\ref{eq13}). 

\section{Concluding remarks}
Summing up, what is the elvisebrion hypothesis about? It suggests a possible
existence of a hidden sector which is either not Lorentz invariant or it is
Lorentz invariant but with a different limiting speed. If the two sectors
(hidden and visible) are connected very weakly, we can encounter a situation 
in which the Lorentz invariance is a very good approximation in the visible 
sector but nevertheless the whole setup is no longer Lorentz invariant and, 
therefore, a possibility of hidden-sector induced superluminal phenomena does 
appear.

On the eve of Higgs discovery, one can speculate about the possibility that
the Higgs field may provide a portal into hidden sectors \cite{69A}.
Usually, the Higgs portal is introduced through the renormalizable coupling 
${\cal L}_{int}=g\Phi^+\Phi\,\Psi^+\Psi$ of the Higgs field $\Phi$ to a 
scalar $\Psi$ from a hidden sector. If the hidden sector is not Lorentz
invariant, such a coupling may induce a contribution to the Higgs mass term
which is space dependent and, therefore, breaks Lorentz invariance
in the Higgs sector. This violation of Lorentz invariance is then propagated 
at one loop level to all other Standard Model particles. Such a scenario was 
considered in \cite{69B}.  

However, if the hidden sector is considered to be not Lorentz invariant, it
is not necessary to assume that its couplings to the visible sector are
Lorentz invariant either. To illustrate the elvisebrion idea, we can consider, 
for example, the following minimal model given by the lagrangian ${\cal L}=
{\cal L}_1+{\cal L}_2+{\cal L}_{int}$, where
\begin{equation}
{\cal L}_1=\frac{1}{c_1^2}\,\frac{\partial \Phi^+}{\partial t}
\frac{\partial \Phi}{\partial t}-\nabla\Phi^+\cdot\nabla\Phi+\mu^2 \Phi^+
\Phi-\frac{1}{2}\lambda(\Phi^+\Phi)^2
\label{eq86}
\end{equation}
is the lagrangian for the Standard Model Higgs doublet $\Phi$,
\begin{equation}
{\cal L}_2=\frac{1}{2}\left[\frac{1}{c_2^2}\,\frac{\partial \Psi}{\partial t}
\frac{\partial \Psi}{\partial t}-\nabla\Psi\cdot\nabla\Psi\right ]
\label{eq87}
\end{equation}
is the lagrangian for the hidden sector presented here (rather 
unrealistically) by only one massless real scalar field $\Psi$ with limiting 
velocity $c_2$ which we assume to exceed the light velocity $c_1$, and
\begin{equation}
{\cal L}_{int}=g\,\Phi^+\Phi\,\frac{\partial \Psi}{\partial x}
\label{eq88}
\end{equation}
is the interaction Lagrangian. Note that ${\cal L}_{int}$ breaks not only 
the Lorentz symmetry, but also the spatial isotropy as it assumes the 
existence of a preferred direction taken here to be the $x$-direction.

Classical equations of motion that follow from such a Lagrangian are
\begin{eqnarray} &&
\frac{1}{c_1^2}\frac{\partial^2 \Phi}{\partial t^2}-\Delta\Phi=\mu^2\Phi-
\lambda(\Phi^+\Phi)\Phi+g\,\Phi\frac{\partial \Psi}{\partial x},
\nonumber \\ &&
\frac{1}{c_2^2}\frac{\partial^2 \Psi}{\partial t^2}-\Delta\Psi=
-g\,\frac{\partial}{\partial x}(\Phi^+\Phi).
\label{eq89}
\end{eqnarray}
As we see, we have a system of coupled nonlinear Klein-Gordon equations with 
different speeds. It will be convenient to introduce dimensionless variables
\begin{equation}
\tilde t=c_1\,\mu\,t,\;\;\;\tilde x=\frac{c_1}{c_2}\,\mu\,x,\;\;\;
\Phi=\frac{\mu}{\sqrt{\lambda}}\,\phi,\;\;\;
\Psi=\frac{c_2}{c_1}\,\frac{\mu}{\sqrt{\lambda}}\,\psi.
\label{eq90}
\end{equation}
Then the system (\ref{eq89}) takes the form
\begin{eqnarray} &&
\frac{\partial^2 \phi}{\partial \tilde t^2}-\epsilon^2\,\tilde \Delta\phi=
\phi-(\phi^+\phi)\phi+\frac{g}{\sqrt{\lambda}}\,\phi\,\frac{\partial \psi}
{\partial \tilde x}, \nonumber \\ &&
\frac{\partial^2 \psi}{\partial \tilde t^2}-\tilde \Delta\psi=
-\frac{g}{\sqrt{\lambda}}\,\frac{\partial}{\partial \tilde x}(\phi^+\phi),
\label{eq91}
\end{eqnarray}
where $\epsilon=c_1/c_2$. When $g/\sqrt{\lambda}=1$, the system 
(\ref{eq91}) describes the three-dimensional dynamics of a twisted elastic 
rod near its first bifurcation threshold \cite{69C,69D} and has many 
interesting solutions \cite{69C}. The similar solutions do exist even in the 
case of small $g$. For example, it can be checked that we have the following
traveling pulse solution
\begin{equation}
\phi=\left (\begin{array}{cc} 0 \\
A\,\mathrm{sech}{\left (\frac{\tilde x -\nu\tilde t}{l}\right )}\end{array}
\right ),\;\;\;\psi=B\tanh{\left (\frac{\tilde x -\nu\tilde t}{l}\right )},
\label{eq92}
\end{equation}
where
\begin{equation}
A^2=\frac{2(\nu^2-1)}{\nu^2-1+g^2/\lambda},\;\;\;
B=\frac{-2gl/\sqrt{\lambda}}{\nu^2-1+g^2/\lambda},\;\;\;
l=\sqrt{\nu^2-\epsilon^2}.
\label{eq93}
\end{equation}
If we assume $\epsilon<1$, then the solution (\ref{eq92}) does exist provided
$\nu>1$. In dimensionfull quantities, the pulse (\ref{eq92}) moves with the
superluminal velocity $v=\nu\,c_2>c_2$ and has the width
$$L=\frac{lc_2}{\mu\,c_1}=\frac{1}{\mu}\,\sqrt{\frac{v^2}{c_1^2}-1}.$$
The solution (\ref{eq92}) survives in the limit $g=0,\;\epsilon=1$ and
describes a tachyonic pulse of the Higgs field. It is a common consensus
that such tachyonic pulses cannot convay information at superluminal speeds.
Does the situation change when the pulse contains a small admixture of
the hidden scalar $\Psi$ whose limiting velocity $c_2$ exceeds the light
velocity $c_1$? We suspect that the answer is yes but do not know for sure.
This is just one question from many that the elvisebrion hypothesis raises.
For example, how is the gravity modified when the hidden sector violates 
Lorentz invariance? Will some kind of ``mirror gravity'' \cite{69E,69F} 
emerge? At this point we do not know answers to these questions.

An unofficial history of tachyons begins with a brief 1959 paper by Sudarshan
sent to Physical Review \cite{69}. The paper was, however, rejected with a
referee report saying that everything was wrong in the paper. Sudarshan 
requested a second referee and a new report claimed that everything was 
right in the paper but all the results were well known. The culmination
of the story was the report of the third referee saying ``I have read
the manuscript, and the two referee reports. I agree with both of them'' 
\cite{69}. As a result the paper was not published and the official history 
of tachyons begins with another paper \cite{37}. We hope that referees will
be more friendly to the elvisebrion hypothesis and it will not generate
confusing and contradictory reports. But what is our own confidence in the
elvisebrion hypothesis?

Martin Rees once said that he is sufficiently confident about the Multiverse 
to bet his dog's life on it. He was supported by  Andrei Linde who was ready 
to bet his own life, and by Steven Weinberg who had just enough confidence 
in the Multiverse hypothesis to bet the lives of both Andrei Linde and Martin 
Rees's dog \cite{70}. We cannot bet the lives of our pets on the elvisebrion 
hypothesis, but have enough confidence in it to bet the lives of both 
Wigner's friend and Schr\"{o}dinger's cat!  
 
Will special relativity, as a fundamental theory, survive for the next hundred 
years? We are not as certain about this as were several years ago. Nowadays
the Lorentz symmetry is frequently questioned by scientists from various
points of views \cite{71}, but there is still no single reliable 
experimental fact indicating the breakdown of special relativity. 

True superluminal particles, elvisebrions, if found, will indicate that
special relativity does not encompass the whole world of material beings,
but it may still be an extremely good approximation in our sector of the 
world for energies that are not very high (compared, probably, to the Planck 
energy, $E_P\sim 10^{19}~GeV$). Therefore, the impressive experimental 
support for special relativity cannot be used as an argument against 
a possible existence of superluminal particles. History of Nature's 
exploration teaches us that ``her bag of performable tricks'' is full of 
wonders.

It is certain, however, that special relativity will remain a precious diamond
of the twentieth century physics. Future developments can only place it in
the proper framework of more general and powerful theory, emphasizing its
sparkling beauty.  

\section*{Note added}
After the main part of this work was completed, we became aware that similar 
ideas had been formulated by Gonzalez-Mestres (see \cite{72,73} and 
references therein). He considers a hypothetical situation when the 
excitations of the ``substrate'' are also governed by special relativity 
(effective or fundamental) with the invariant speed $c_s$ which is much larger 
than the light velocity, the invariant speed  of the effective relativity
realized in our sector of the world. Interestingly, there exists a ready
condensed matter analogy, albeit two-dimensional, of such a situation. 
In graphene, the low energy electronic states are described by the Dirac 
equation for zero-mass particles \cite{74} and an effective relativity emerges
with the Fermi velocity, $v_F\sim 10^6~m/s$, in the role of invariant 
velocity. The Fermi velocity in graphene is much smaller than the velocity of
light. Therefore, cosmic ray particles which traverse a graphene sheet
will appear as elvisebrions (Gonzalez-Mestres calls them superbradyons) in the
world of graphene electrons, and their velocity can easily exceed $v_F$.

After this article was completed, we became aware of the very interesting 
paper \cite{75} by Geroch which presents other arguments explaining why 
elvisebrions could exist without any conflict with the well-established and 
overwhelming experimental evidence of relativity. 

Let us also mention an interesting contribution by Unzicker \cite{76}. He
reconsiders the compatibility of the concept of the aether with special 
relativity and concludes that ``not the concept of the aether as such is 
wrong, but the idea of particles consisting of external material passing
through the aether. Rather the aether is a concept that yields special 
relativity in a quite natural way, provided that topological defects are 
seen as particles'' \cite{76}. Although he does not considers elvisebrions, 
from such a picture (relativistic particles as topological defects of the 
aether) there is just one step to assume a possibility of coexistence both of
topological defects and of external particles passing through the aether 
(elvisebrions).

\section*{Acknowledgements}
The work of Z.K.S. is supported by the Ministry of Education and Science of 
the Russian Federation and in part by Russian Federation President
Grant for the support of scientific schools NSh-5320.2012.2.

\end{document}